\pdfoutput=1
\documentclass[acmlarge]{acmart}

\usepackage{booktabs} 
\usepackage{graphicx}
\usepackage{epstopdf}
\usepackage{subfig}
\usepackage{url}
\usepackage{kotex}
\usepackage{color}
\usepackage{verbatim}
\usepackage{hhline}
\usepackage{here}
\usepackage{multicol}

\usepackage{algpseudocode}

\usepackage[toc,page]{appendix}

\usepackage{algorithm}

\usepackage{algcompatible}

\newsavebox{\algleft}
\newsavebox{\algright}

\newcommand{\specialcell}[2][c]{%
  \begin{tabular}[#1]{@{}c@{}}#2\end{tabular}}

\newcommand{\BREAK}{\STATE \textbf{break} }

\acmJournal{IMWUT}
\acmVolume{0}
\acmNumber{0}
\acmArticle{0}
\acmYear{2017}
\acmMonth{5}
\acmArticleSeq{0}

\setcopyright{usgovmixed}

\acmDOI{0000001.0000001}

\received{May 2017}

\begin{document}
\title{QuickTalk: An Association-Free Communication Method for IoT Devices in Proximity}

\author{Seongmin Ham}
\affiliation{%
  \institution{School of ECE, UNIST}
  \city{Ulsan} 
  \country{Republic of Korea}
}
\email{smham@unist.ac.kr}

\author{Jihyung Lee}
\affiliation{%
  \institution{School of ECE, UNIST}
  \city{Ulsan} 
  \country{Republic of Korea}
}
\email{im21cace@unist.ac.kr}

\author{Kyunghan Lee}
\affiliation{%
  \institution{School of ECE, UNIST}
  \city{Ulsan} 
  \country{Republic of Korea}
}
\email{khlee@unist.ac.kr}

\begin{abstract}
IoT devices are in general considered to be straightforward to use. However, we find that there are a number of situations where the usability becomes poor. The situations include but not limited to the followings: 1) when initializing an IoT device, 2) when trying to control an IoT device which is initialized and registered by another person, and 3) when trying to control an IoT device out of many of the same type. We tackle these situations by proposing a new association-free communication method, QuickTalk. QuickTalk lets a user device such as a smartphone pinpoint and activate an IoT device with the help of an IR transmitter and communicate with the pinpointed IoT device through the broadcast channel of WiFi. By the nature of its association-free communication, QuickTalk allows a user device to immediately give a command to a specific IoT device in proximity even when the IoT device is uninitialized, unregistered to the control interface of the user, or registered but being physically confused with others. Our experiments of QuickTalk implemented on Raspberry Pi 2 devices show that the end-to-end delay of QuickTalk is upper bounded by 2.5 seconds and its median is only about 0.74 seconds. We further confirm that even when an IoT device has ongoing data sessions, QuickTalk can still establish a reliable communication channel to the IoT device with little impact to the ongoing sessions.

\end{abstract}




\maketitle

\renewcommand{\shortauthors}{S. Ham et al.}

\section{INTRODUCTION}
In the last few years, IoT (Internet of Things) has been arguably one of the most commercially promoted technical terms in the field of computer networks. Comparing to its early stage where IoT was just a concept of connecting numerous small devices such as sensors, actuators, and embedded systems to the Internet, it is now much more mature with a number of in-situ realizations. Such realizations that are often found in the area of home automation include thermal controllers~\cite{NEST}, wattage monitors~\cite{NEURIO}, gas valves~\cite{LGGAS}, and lighting adjusters~\cite{HUE}. These IoT devices are distinguished from their conventional forms by having not only the ubiquitous accessibility but also the software interface that guarantees virtually the same or even improved usability compared to local physical transactions. 
Thanks to these properties, IoT devices are considered to be convenient and easy to use. 


\begin{figure*}[t!]
    \centering
    \subfloat[Alice's parent can control the IoT device through the control interface.\label{fig:a}]{\includegraphics[height=7cm]{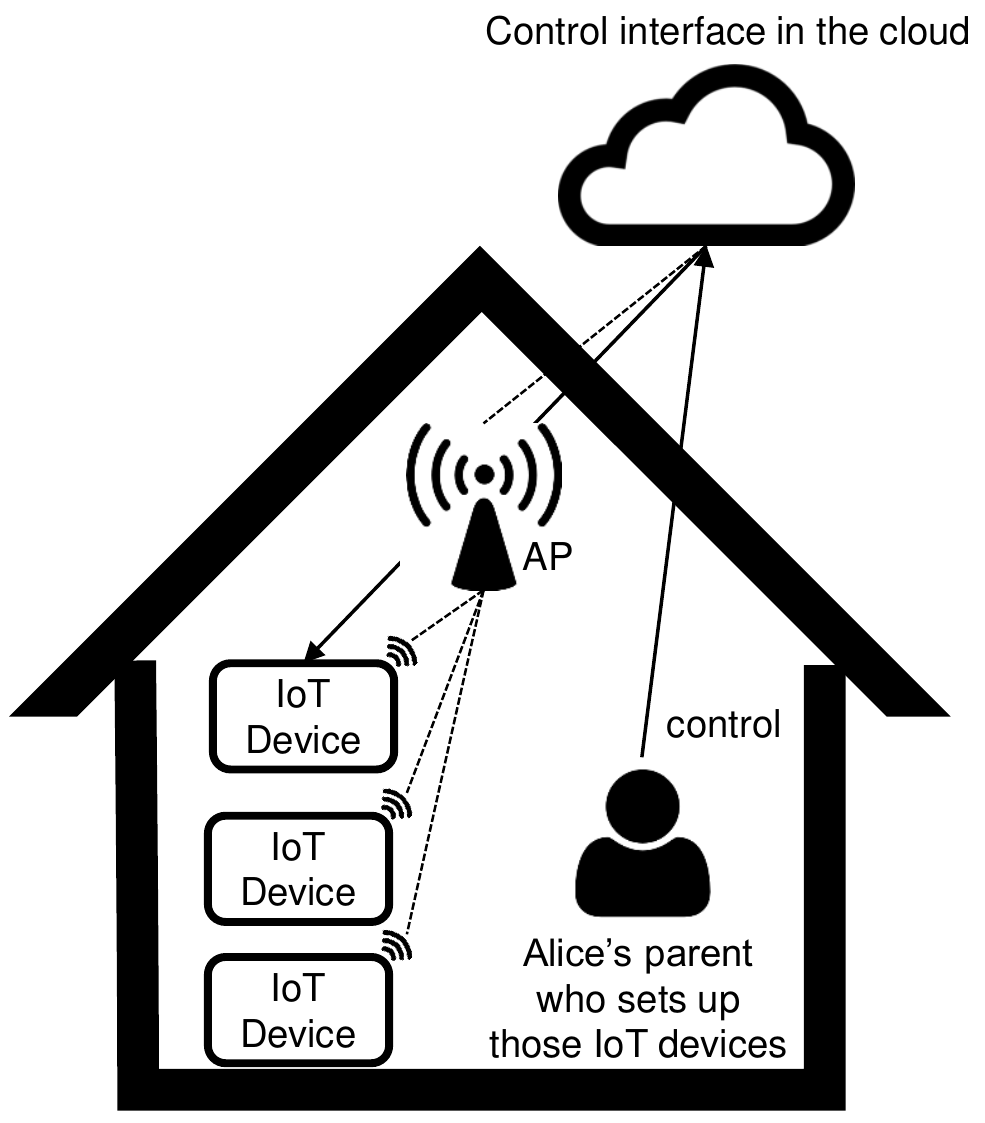}}\qquad
    \subfloat[Alice cannot control the IoT device. Even she is located close to the IoT device. \label{fig:b}]{\includegraphics[height=7cm]{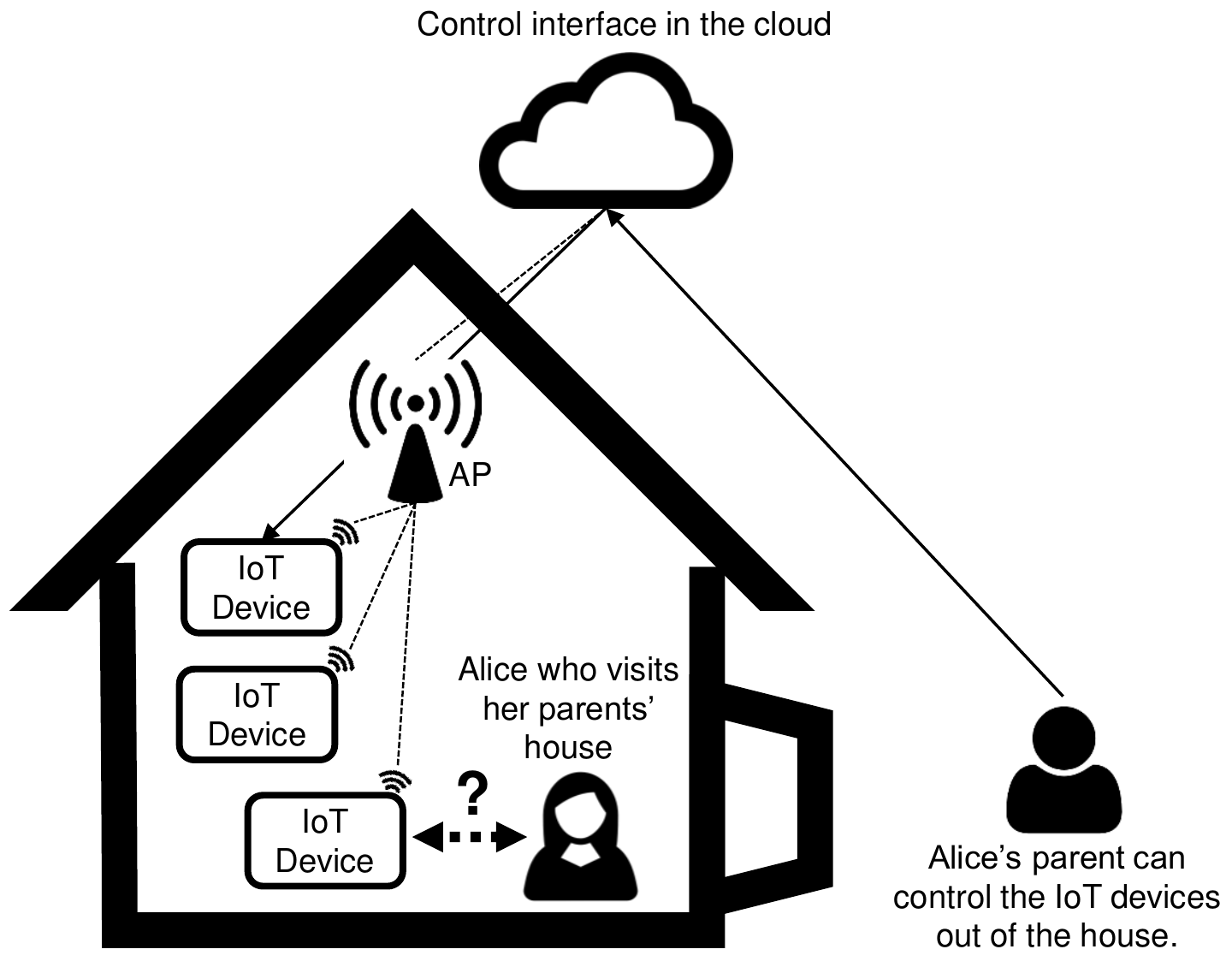}}
    \caption{\textcolor{black}{Controlling the IoT devices in proximity is not always straightforward. There exist situations where some visitors are naturally allowed to control devices in the proximity, but the current IoT devices are not flexible enough to accommodate such users.}}
    \label{fig:concept}
\end{figure*}

\textcolor{black}{
However, we find that there are critical situations where the usability of IoT devices is disappointing. 
These problems are revealed by the following use cases: a) Alice visited her parents' house and found that there exists an IoT thermal controller. She wanted to change the setting of that controller but noticed that she cannot do anything without having the smartphone of her parent, which previously set up the device. \textcolor{black}{This scenario is depicted in Fig.~\ref{fig:concept}}, 
b) Bob who manages an office installed a bunch of IoT window shades that can regulate the amount of sunlight by the settings. He originally programmed them in a group to react to the indoor temperature but soon identified that a few of them need adjustment. However, he soon noticed that controlling an individual window shade is not possible when its identifier is not known. He had to spend a substantial amount of time to manually verify those identifiers and wondered why an IoT device is not directly controllable although it is within his sight, 
and c) Charlie who wants to publicize information such as air quality and traffic situation through outdoor IoT sensors planned to distribute this information to the passing-by users. But he realized that there is no way for the users to get this information unless they come close and capture a QR code~\cite{QRcode} or an NFC tag~\cite{NFC}, either of which is connected to a web page or an application. }

\textcolor{black}{
Situation a), what we call by \textit{visitor problem}, points out that the current IoT devices are designed to be controlled only by the owner who set up the devices through an application (e.g., iHome) or an web interface (e.g., Google The Physical Web). Therefore, unless a visitor borrows the control device having the application or the identification information (i.e., ID and password) to the web interface, it is impossible for the visitor to control any device that is visible in the proximity. This is counter-intuitive to most non-tech savvy users.}

\textcolor{black}{
Situation b), what we call by \textit{naming problem}, brings up a challenge in the setup process. When there are only few IoT devices of the same type in a place, naming is not an issue. However, as shown in the scenario, if a user has to control a set of devices, for instance, a hundred, which are installed closely to each other, a typical naming scheme such as \textit{shade\_1} or \textit{shade\_meeting-room} no longer works. Given the future of IoT environments where dense deployments would prevail, the user inconvenience involved in controlling the devices after judiciously configuring their names would soon become more prominent.}


\textcolor{black}{Situation c), what we call by \textit{range problem}, pulls out an issue of access distance toward an identifier of various forms. Because the size of a QR code is limited by the design purpose and the communication range of NFC is designed to be in the order of centimeters to avoid any misconnection, making the first connection to an IoT device in some distance (e.g., few tens of meters) is currently not.}

From the aforementioned problem statements, we observe that these problems root from a single cause, that is a lack of a feature which allows a user device to communicate with a specific IoT device without going through an association process. We call this feature \textit{association-free communication} for the IoT devices in proximity. Once this feature is enabled, a user no longer suffers from the exemplified situations. However implementing the feature brings new technical challenges: 1) how to pinpoint a device in proximity and 2) how to set up a communication channel without an association process \textcolor{black}{while preserving existing sessions if there is any}. We tackle these challenges by proposing \textit{QuickTalk} that uniquely combines IR (Infrared) signal emission and WiFi overhearing over the broadcast channel~\footnote{\textcolor{black}{QuickTalk specifically exploits WiFi in its implementation since it is the most widely deployed wireless technique that is both available in smartphones and IoT devices. However, note that Quicktalk is not limited to work with WiFi and other low-power wireless techniques such as Zigbee and BLE can also be adopted.}}. 

In a nutshell, QuickTalk at a user device utilizes IR to pinpoint and trigger an IoT device and to deliver the ID of the user device (e.g., WiFi MAC address). Upon reception of the ID, the IoT device keeps broadcasting the ID through its WiFi interface to its current WiFi channel so that the user device can detect the channel of the IoT device by extracting the ID while sweeping the WiFi channels. Once the channel is known, QuickTalk lets them communicate with each other by WiFi broadcasts at that channel.  

We implement QuickTalk as software stacks for Raspberry Pi 2 devices that emulate a user device and an IoT device with IR and WiFi interfaces. Our validation reveals that thanks to the strong directivity nature of IR, QuickTalk can pinpoint and trigger an IoT device almost immediately with a narrow-angle of $\pm$10 degrees and also thanks to the nature of broadcast communication, QuickTalk allows ongoing communications at the IoT device, if any, to coexist with the newly established broadcast communication. These advantages make QuickTalk to be applicable to the IoT devices that are fresh out of the box and even to the IoT devices that are densely deployed. 



\section{RELATED WORK}
There exist a huge number of studies in the context of designing IoT systems. In this section, instead of providing a broad introduction, we give our focus to the previous studies that are highly relevant to our work in the following two aspects: 1) networking architecture and 2) control interface for IoT devices. 

\subsection{Networking Architecture}
Early-stage IoT devices that showed little difference to the nodes of sensor networks were mostly relying on synchronization-based networking methods, where each device accumulates its sensor readings and periodically synchronizes the readings to a local or a remote server in batches~\cite{atzori2010internet}. Guinard et al.~\cite{guinard2009towards} pointed out the networking inefficiency (e.g., overhead, data freshness) of using synchronized-based methods in IoT systems and proposed a resource-oriented networking architecture which conforms to the principles of REST (Representational State Transfer) and utilizes embedded HTTP (Hypertext Transfer Protocol). 

Later, IETF 6LoWPAN working group~\cite{bormann} raised performance issues of using HTTP or other TCP-based protocols in the typical environment of operating IoT systems, which is highly constrained mainly due to the instability of network links, limited computing capability, and relatively small battery capacity of IoT devices.

To alleviate those issues, a light-weight protocol, CoAP (Constrained Application Protocol) was proposed~\cite{shelby} and then was standardized by IETF (Internet Engineering Task Force) as RFC 7252~\cite{CoAP}. An experimental study by Leva et al.~\cite{leva2014comparing} showed that CoAP is indeed light-weight by demonstrating that an IoT device can save about 70\% of maintenance cost of the battery when using CoAP compared to using HTTP with a total cost of ownership model. To achieve its goal, CoAP is designed to use UDP (User Datagram Protocol), to be RESTful, and to be easily translatable to HTTP. Although CoAP allows direct communication between IoT devices and their user devices, RFC 7252 suggests using CoAP mainly between the IoT devices and an HTTP proxy that serves as a gateway or a control hub for the user devices. Because an HTTP proxy and user devices can communicate with HTTP in a regular manner, when an HTTP proxy exists, it is not essential for the user devices to understand CoAP. An HTTP proxy can be typically placed in the same network where IoT devices belong to but it is possible for the proxy to be placed in a cloud platform to enable ubiquitous accessibility toward IoT devices. Placing an HTTP proxy in a cloud is proposed by Kovatsch et al.~\cite{kovatsch2012moving} and this idea is now implemented in various commercial cloud platforms such as Microsoft Azure~\cite{MSAzure}, Amazon Web Service~\cite{AWS}, and Apple iCloud~\cite{iCloud}. 

\subsection{Making Connections to the Devices in the Proximity}
\textcolor{black}{To enable communication between a user device (e.g., smartphone, remote controller) and another device in the proximity, there have been various connection automation methods exploiting wireless signal exchange techniques such as IR~\cite{kindberg2001web}, RFID ~\cite{want1999bridging}, and NFC~\cite{NFC}, as summarized in the survey from Edwards et al.~\cite{edwards2006discovery}. Although those methods rely on different technologies, they essentially share the same idea that consists of two step processes: 1) designating a single device to make a connection by sufficiently limiting the range or the angle of wireless communication, and 2) obtaining the WiFi or Bluetooth credential information of that device to secure a longer range communication channel. This idea is practical and is current being supported by diverse home appliances such as speakers and televisions. However, as the devices in the proximity become denser in both commercial and personal environment, a new demand for distinghishing devices in the communication range has emerged. To this end, Zhang et al.~\cite{zhang2014hobs} proposes an IR pointing accuracy improvement technique that exploits the ordering of the received signal strengths at multiple IR receivers. Xie et al.~\cite{xie2016tell} suggests to combine a depth camera and a rotating antenna-array to locate multiple RFIDs that are visible at the camera (i.e., observable by the user of the camera). Kong et al.~\cite{kong2016selecting} takes a more radical approach of visually recognizing an object by a machine learning technique so as to find the matching device from the list of devices in the proximity. QuickTalk adopts a few nice features from these techniques (e.g., IR accuracy improvement), but it is still unique in the sense that it does not disrupt on-going communication sessions of the target device if there exists any. Thanks to this uniqueness, QuickTalk is the only mechanism that allows multiple users to make simultaneous connections to a single device without a monopoly.}

\section{SYSTEM DESIGN}
In this section, we first provide a problem statement that we are trying to solve through this work. Then, with an overview of the software architecture of QuickTalk, we explain the required function blocks of a user device and an IoT device that use QuickTalk. Lastly, we introduce technical challenges involved in enabling QuickTalk in a mobile device.    


\begin{figure}[t]
	\centering
	\includegraphics[width=0.7\columnwidth]{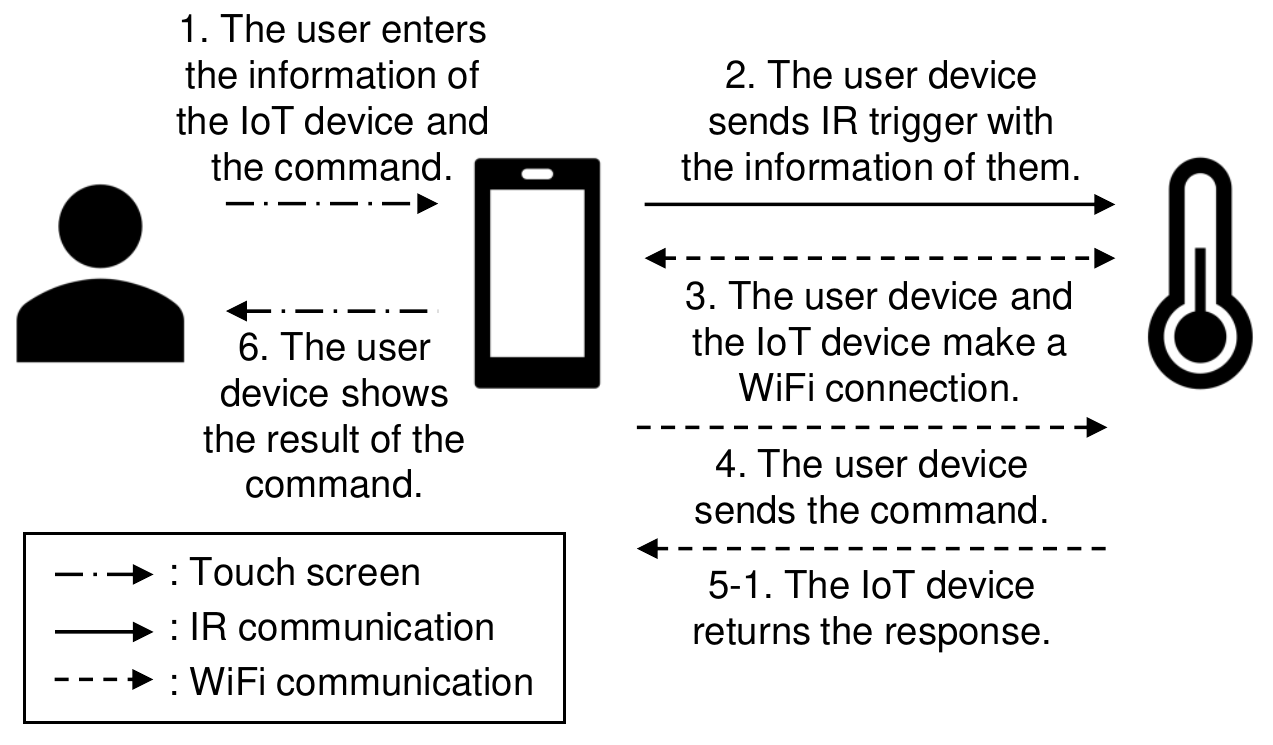}
	\caption{\textcolor{black}{Overview of the QuickTalk.}}
    \label{fig:overview}
\end{figure}
\subsection{Problem Statement}
Regarding the initialization of an IoT device, an IoT device is in either of two states: 1) \textit{uninitialized} and 2) \textit{registered}. Uninitialized state of an IoT device means that the device is just taken out of its box and it is not currently connected to any control interface. Every IoT device is given to a user at its uninitialized state and requires the user to go through a certain setup procedure. Once the setup procedure is completed, the state of an IoT device changes to registered and becomes controllable by the associated control interface. When becoming registered, IoT devices in the market are mostly connected through WiFi interface allowing NLOS(Non-Line-of-Sight) controls. It is important to note that once an IoT device is registered to a control interface, its control permission is given to the person or the group of persons who own the access permission to the control interface. Thus, users who are not reachable to the control interface of an IoT device, are prevented from controlling the IoT device. 
\textcolor{black}{
The above-mentioned method of controlling IoT devices through a control interface is reasonable in general, but gives challenges to an IoT novice when she is visiting a new place and is seeing many IoT devices in the place that are open to any visitor. One way of guiding her to use those IoT devices is to authorize her to access the control interface in which all those IoT devices are pre-registered. However, when the place has a large number of visitors, authorizing individual visitors is not secure given that the control interface is accessible from anywhere through the Internet once authorized. Invalidating the authorization when she leaves the place is possible but is of too much complication. Also, when there are too many IoT devices in the place, it is difficult for her and other visitors to identify which entity in the control interface corresponds to which actual IoT device. }

As a solution to these matters, our goal is to provide an immediate and intuitive method of controlling IoT devices to a user who is in the proximity of those IoT devices. We aim at designing a new method that is able to physically pinpoint an IoT device and to any IoT device that is either uninitialized or registered. After that, a short moment of pinpointing, we also aim at controlling the IoT device at any direction as long as we stay in the proximity.

\subsection{Candidates for Pinpointing} 
\textcolor{black}{ 
As explained in the previous section, we have several candidates pinpointing methods: IR, QR code, RFID, NFC, audio signal, and image processing. We conclude that IR is the only viable solution by the following constraints: 1) The pinpointing distance should be at least a few meters and it is great if the distance can be extended to a few tens of meters, 2) The effort of a user to pinpoint should be minimal, and 3) The possibility of ambiguity for a pinpointing action should be minimal (i.e., the accuracy of pinpointing should be maximal). NFC, QR code, and short-range RFID are not qualified by the first requirement, and long-range RFID and image processing fail to meet the third requirement. Audio signal that relies on the Doppler shift violates the second requirement.}

\textcolor{black}{
IR communication is sometimes misunderstood as if it is obsolete, however, this is not true. There are still various companies such as Huawei, LG, and Xiaomi that support IR in their smartphones. Google also began to support native IR blaster in Android OS from API version 19~\cite{Kitkat} since 2013. On the other hand, it is true that most IoT devices do not have an IR receiver recognizing that it is being pointed. However, given its usefulness and low cost, installing an IR receiver in IoT devices is a viable option.
}



\subsection{QuickTalk Architecture}

\textcolor{black}{
We design the workflow of QuickTalk as it is depicted in Fig.~\ref{fig:overview}. To enable QuickTalk between a user device and an IoT device, QuickTalk implements two services: IR service for pinpointing and WiFi service for communication. As it is explained in the problem statement, it is natural for users to control IoT devices from any direction as long as they stay in the proximity. Therefore, to avoid any LOS-based communication through IR, the IR service at a user device is designed to only shoot the beam and the IR receiver at the designated IoT device is designed to only capture the beam. When shooting and detecting the beam, both IR transmitter and receiver respectively trigger their WiFi services and make a connection. More detailed specifications of a user device and an IoT device are provided below.}

\subsubsection{User Device}
For an immediate control of an IoT device from a user device, the user device for QuickTalk utilizes a concept of the filter, called device-type filter, and asks a user to provide the type of the IoT device to be controlled when emitting a command to that IoT device. The filter is also helpful to significantly reduce the possibility of experiencing confusion in pointing an IoT device, especially when there are many IoT devices in close proximity. For instance, when trying to control an IoT bulb out of many IoT devices in the same place, a user of QuickTalk may point to the bulb with the filter specified as \textit{BULB} to maximize the pinpointing efficiency. Once a user device succeeds in specifying and activating an IoT device within a very short time, QuickTalk switches to utilize WiFi instead of IR for the reliable (i.e., NLOS) delivery of user commands to the IoT device. We will discuss the challenges involved in switching to WiFi in the next subsection. 

\subsubsection{IoT Device} 
An IoT device with an IR receiver reacts when an IR signal is detected at its IR receiver. As will be shown later, it is possible to enable QuickTalk without having an IR receiver in an IoT device. However, having an IR receiver which is of low-cost adds two major benefits to an IoT device: 1) intuitive pointing (i.e., specification) and 2) energy efficiency. By the nature of strong directionality of IR signal, when IR transmitter and receiver are properly installed, a very narrow pinpointing ability is achievable. We will discuss the pinpointing ability of IR signal in the following section. Regarding the energy efficiency, an IR receiver does an important role. Suppose that there is an uninitialized IoT device or an IoT device that is registered but in a power saving mode. In both cases without having an always-on low-power channel that immediately activates the WiFi interface, a user who wants to deliver a command should wait until the WiFi interface becomes active (e.g., until the duty cycle of power saving ends). Thus, installing a small IR receiver in an IoT device that consumes only less than 10mW is a reasonable choice to achieve energy efficiency while keeping the property of an immediate control. To improve the energy efficiency, QuickTalk utilizes the device-type filter that arrives at an IoT device inside the pointing IR signal, and selectively activates the IoT device of the matching type (e.g., a \textit{THERMAL Controller} on the same line of sight or IR signal is not activated when a device type is specified as \textit{POWER PLUG}).

\subsection{Technical Challenges}
There are two major technical challenges that need to be tackled to enable QuickTalk. The first is to reveal the pinpointing ability of IR signal in real IoT environments and to design the frame structure of the IR signal for reliable message delivery. The second is to design and implement a WiFi communication method that can exchange packets either with uninitialized IoT devices or registered IoT devices.

\subsubsection{IR Pinpointing} 
The most important feature of the IR signal to be studied for QuickTalk is its ability of pinpointing an IoT device. Given that most remote controllers using IR work well in an indoor situation, our focus is given more onto the pinpointing ability at an outdoor environment. Also, the reachable distance and the allowed amount of slanted angle for pinpointing are also of our interest. In addition to the study of these properties, engineering decisions related to the channel coding of the information bits in an IR transmission, the types of information bits to be included, and the length of information bits for an IR transmission are all to be discussed and designed in detail. We provide our answers to these matters in the next section. 

\subsubsection{Association-Free WiFi Communication}  
\textcolor{black}
{In order to make QuickTalk be an immediate control method that works universally to any IoT device, a WiFi communication method that works a user device with an IoT device is essential. Skipping the association process is not only convenient but also powerful as it can allow an IoT device with an ongoing session to communicate with a user device without interrupting the existing session. We call such a feature as \textit{association-free WiFi communication}. Our association-free communication whose detailed mechanism will be explained in the following section does not require authentication (i.e., access control with security protection), because the security concern is resolved in advance by the IR pinpointing process.
}

\section{PROPOSED METHODS}
\label{sec:solution}

\begin{figure}[t!]
	\centering
	\includegraphics[width=0.9\columnwidth]{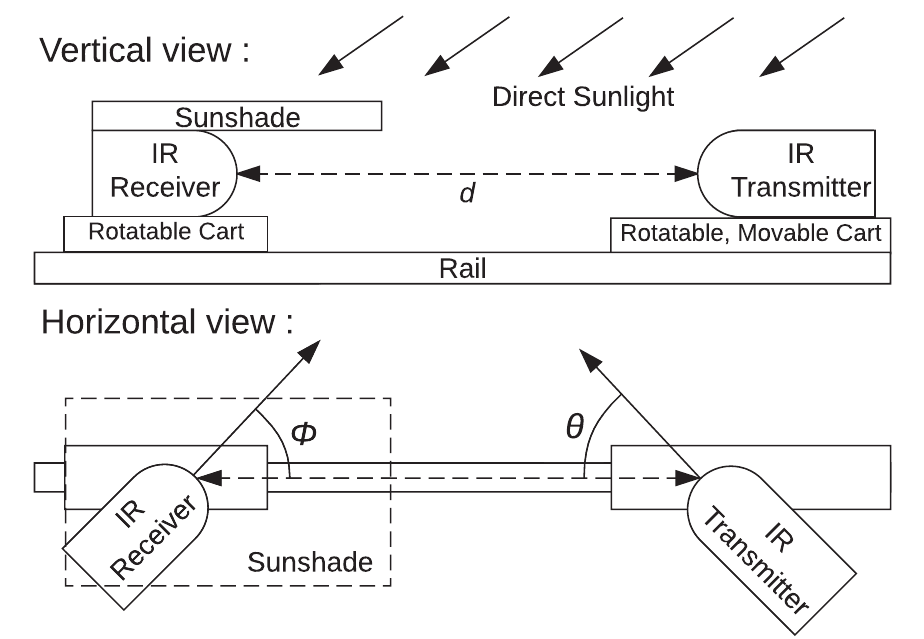}
	\caption{Vertical and horizontal views of the test platform. The distance and angles are controllable as the IR transmitter is on a rotatable and movable cart and the IR receiver is on a rotational cart.}
	\label{fig:testplatform}
\end{figure}

\begin{figure}[t!]
    \centering
    \subfloat[Indoor: Transmission angle ($\theta$) varies while $\phi = 0$.]
    {
        \includegraphics[width=0.9\columnwidth]{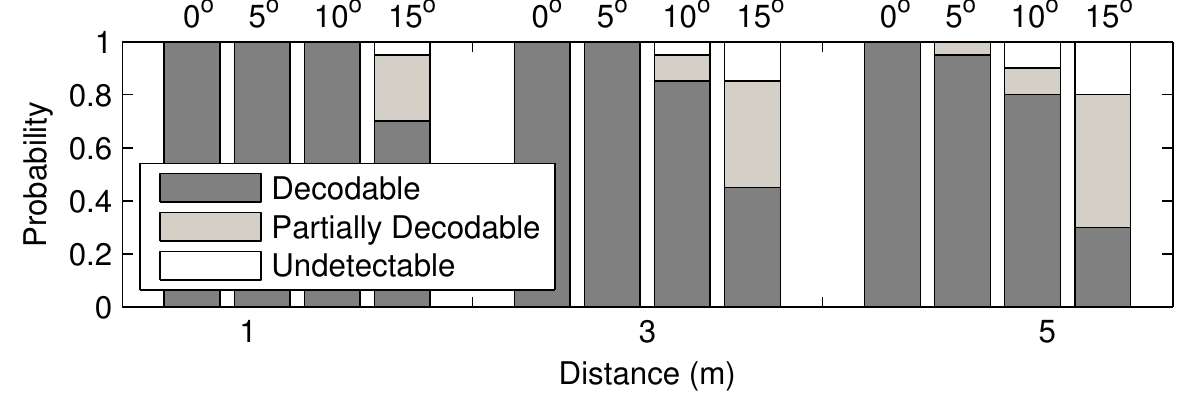}
    }
    \\
    \subfloat[Indoor: Reception angle ($\phi$) varies while $\theta = 0$.]
    {
        \includegraphics[width=0.9\columnwidth]{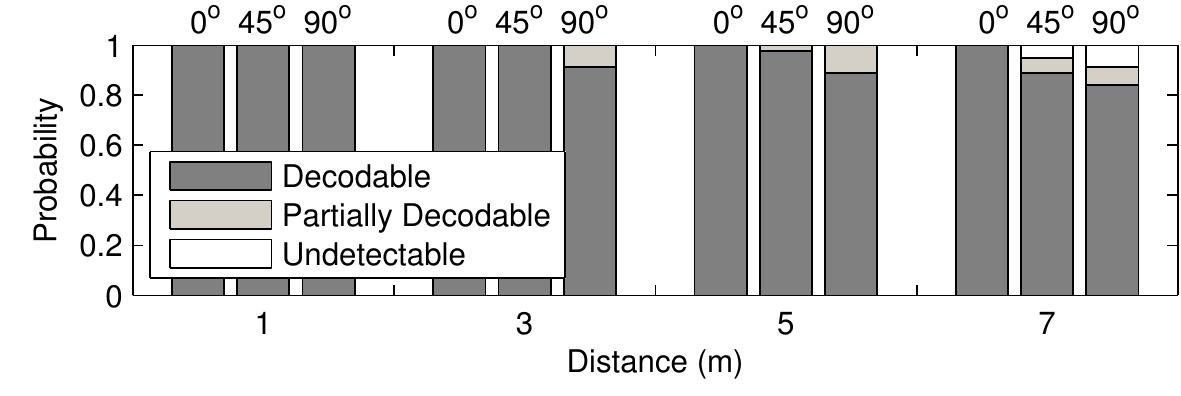}
    }
    \\
    \subfloat[Outdoor: The case of aligned angles, $\theta = \phi = 0$. ]
    {
        \includegraphics[width=0.9\columnwidth]{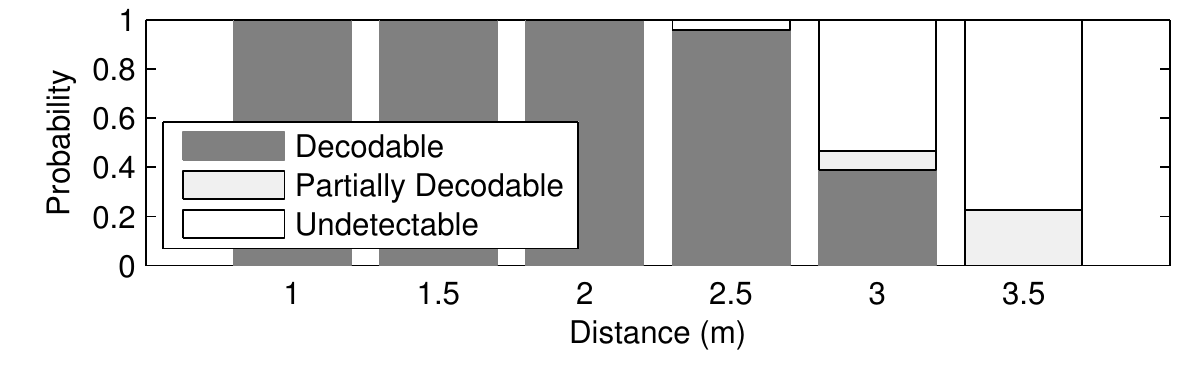}
    }
    \caption{The probability distribution of the received IR signal over the cases: decodable, partially decodable, and undetectable, where the IR signal exchange is experimented at indoor with varying (a) transmission angle ($\theta$), (b) reception angle ($\phi$), and (c) at outdoor with aligned angles.}
    
    \label{fig:IRexperiment}
\end{figure}

In this section, we propose our systems designs as solutions to the challenges raised in implementing IR pinpointing and association-free communication, and validate our proposed designs using Raspberry Pi version 2 devices with 4.1.19 Linux kernel installed, which emulate both user devices and IoT devices. 

    

\subsection{IR Pinpointing}
The purpose of IR pinpointing is twofold: 1) to specify an IoT device to control and 2) to trigger the WiFi interface hence enabling association-free communication. In order to achieve both, we design the data frame emitted whenever a user device pinpoints an IoT device. 
Our design is to emit the ID of the user device in the form of its WiFi MAC address (24bits) and emit device type filters in a hierarchical manner (total 14 bits that are divided into 4, 4, and 6 bits by the levels of categorization) along with parity bits (2bits). In total, each IR signal emission delivers 40 bits of information to an IoT device, which typically takes less than 95 milliseconds. The reason why we have hierarchical device type filters is to maximally narrow down the pointed candidates, hence finally pinpointing a device. For instance, an interactive IoT advertisement display can be hierarchically classified as \textit{DISPLAY:AD-DISPLAY:INTERACTIVE-AD-DISPLAY}. Using the concept of the hierarchical filter, a user device may trade off the user convenience (e.g., user knowledge) of pinpointing and the precision of pinpointing, which is out of the scope of this work and will be of a separate study.

\subsubsection{Validation}
Using one of a de-facto standard of IR communication, called NEC format~\cite{NEC}, we validate if our data frame delivery through IR can pinpoint an IoT device. For the validation, we made a test platform as described in Fig.~\ref{fig:testplatform}. As it is shown, it is to easily test various factors such as transmission distance~\textcolor{black}{(\textit{$d$})}, transmission angle~\textcolor{black}{(\textit{$\phi$})}, and reception angle~\textcolor{black}{(\textit{$\theta$})}. By using the rail and the rotational and movable cart, we performed extensive validations on IR transmitter and receiver that are both operated at the modulation rate of 38 KHz. As the NEC format specifies, our IR signal transmission accompanies 13.5ms\textcolor{black}{(9ms of ON period and 4.5ms of OFF period)} of lead code emission at the beginning for the purpose of separating each IR signal. After the leader code emission, our data frame of 40 bits is followed. Since we find that the repetition of the entire bits followed by the data frame transmission, which is specified in NEC format does not critically affect the success rate of IR transmissions, we intentionally omitted this repetition part for simplicity.

    

Fig.~\ref{fig:IRexperiment} (a), (b), and (c) show that how the data frame sent from a user device is delivered to an IoT device at different settings. We classify the situations by whether the delivered bits are fully decodable, partially decodable, or undetectable. As shown in Fig.~\ref{fig:IRexperiment} (a), it is natural to observe that the longer the distance, the narrower the decodable angle is. However, it is important to note that even at the distance of 5 meters, we can manage to narrow down the decodable angle to be less than 10 degrees, meaning that IoT devices in 5-meter distance will be activated when they are located at a circle of 88.2 centimeters at that distance. Considering the typical spacing of IoT devices, it is fair to say that it is of a sharp pinpointing. According to a recent study~\cite{chen2013context}, the decodable angle of an IR transmitter can be physically controlled without manufacturing a high-cost transmitter only by adjusting the depth of installing an IR transmitter in its housing. Fig.~\ref{fig:IRexperiment} (b) further shows that when it is well pointed by an IR transmitter, the mismatch in the reception angle of an IR receiver does not degrade the decodability. Considering that IoT devices can be installed in various postures, hence having their IR receivers headed toward random directions, our observation in Fig.~\ref{fig:IRexperiment} (b) is optimistic to the users who want to remotely control such IoT devices. Finally, Fig.~\ref{fig:IRexperiment} (c) reveals that IR signal exchange up to 2.5 meters is even possible at a sunny outdoor environment as long as an IR receiver is shaded from the direct sunlight. Overall, we confirm that our IR system design is practically viable in pinpointing an IoT device. 

\subsection{Association-Free WiFi Communication}
To our knowledge, there can be two different ways of implementing a WiFi communication method for QuickTalk, which can coexist with established WiFi sessions if any. 

\begin{figure}[t]
    \centering
    \subfloat[Without retransmission]
    {
	    \includegraphics[width=0.98\columnwidth]{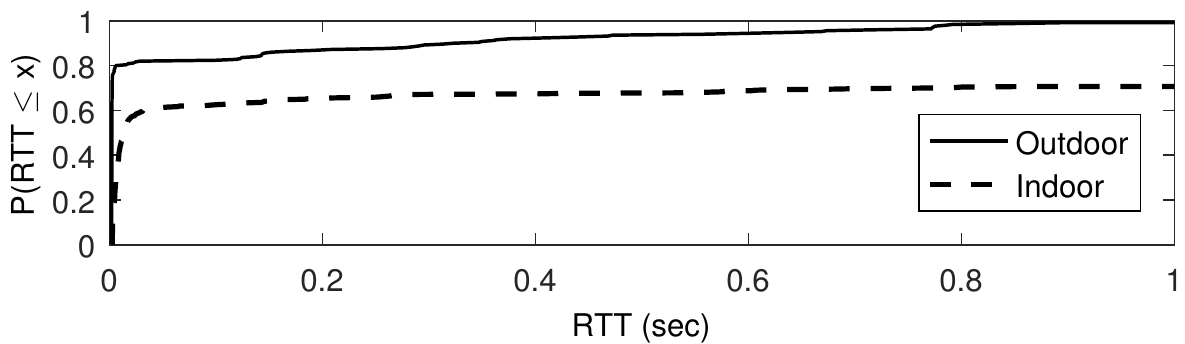}
    }
    \\
    \subfloat[With retransmission]
    {
        \includegraphics[width=0.98\columnwidth]{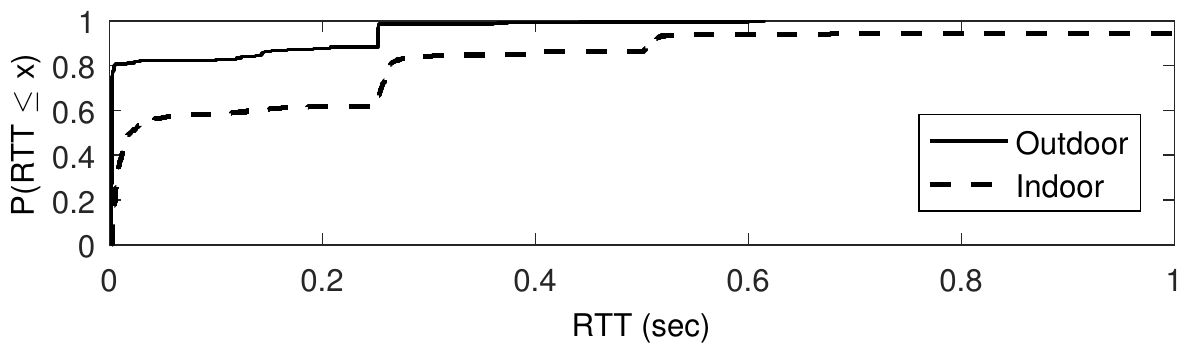}
    }
	\caption{The CDFs of RTT from the association-free communication exploiting packet broadcasts. (a) When there is no application-level retransmission, about 93.6\% and 67.6\% of packets are replied within 0.5 seconds at an outdoor and an indoor environment, respectively. (b) The percentages increase to 99.5\% and 86.5\% when adding the application-level retransmission that retries after 0.25 seconds.} 
    \label{fig:echo}
\end{figure}

The first is to use a \textit{fake PS poll}, where PS poll stands for the power save poll defined in the 802.11 standard~\cite{ieee2005802}. The main idea behind this method is to let an IoT device previously registered to a control interface send a fake PS poll to the access point that the IoT device is being associated upon reception of IR signal. When a fake PS poll is received at the access point, it is known by a work~\cite{gu2008security} that the access point suspends all the ongoing sessions and queue the undelivered packets in the access point. Thus, the IoT device is able to secure a certain amount of period that can be used to communicate with a user device who triggered the IoT device by an IR signal. This is a working method but is a hack, which is not recommended. 

\begin{figure}[t!]
	\centering
	\includegraphics[width=0.85\columnwidth]{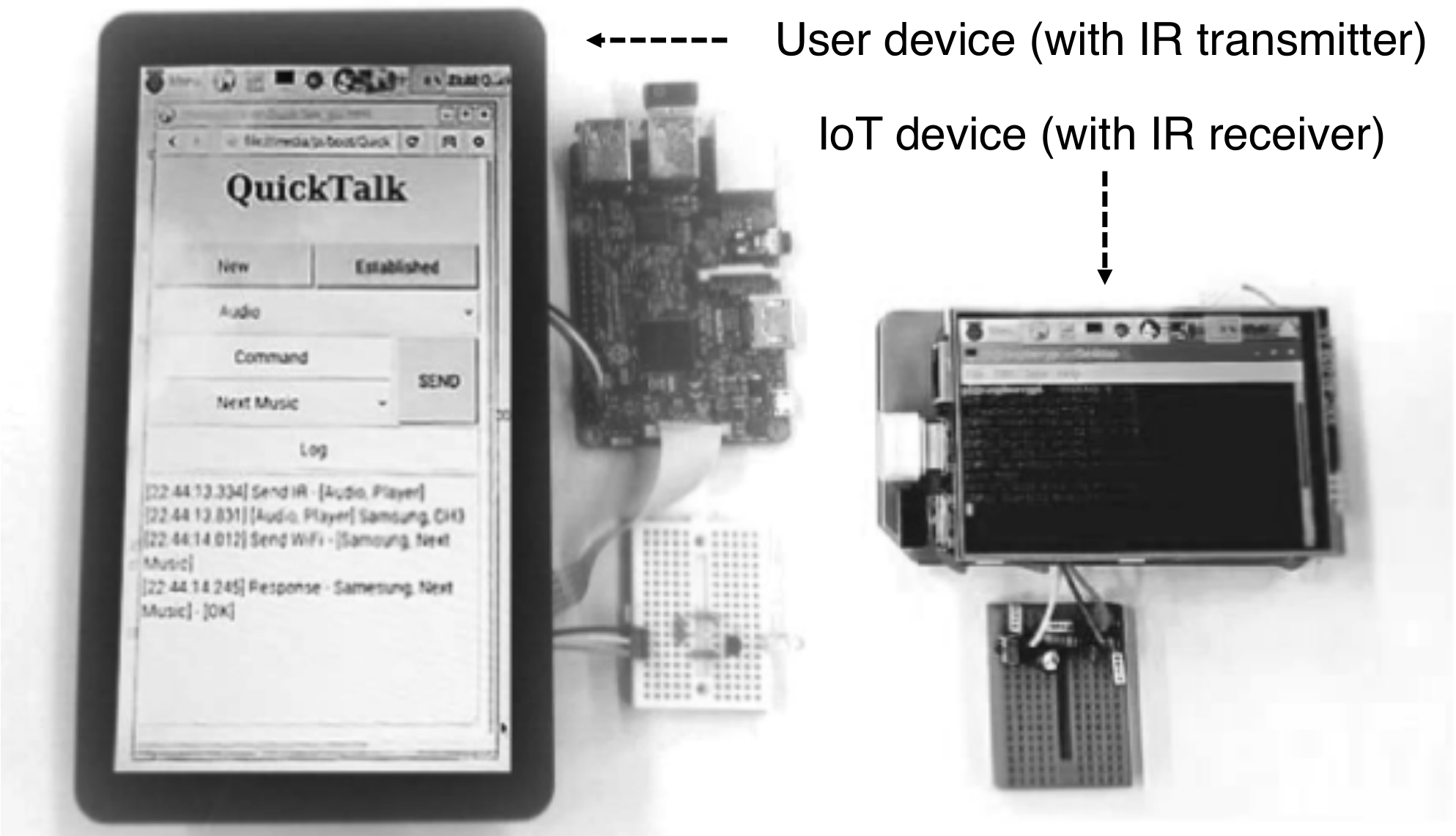} 
	\caption{QuickTalk implementation for a user device (left) and for an IoT device (right). The screen attached to the user device shows our user interface.}
	\label{fig:app}
\end{figure}  

The second is to use the intrinsic packet broadcast ability and the packet monitoring (i.e., capturing) ability of WiFi. We name such a method as \textit{association-free communication}. To enable association-free communication, upon reception of an IR signal, we let the IoT device broadcast the MAC address of the user device which is received through the IR signal and let the user device go to the packet monitor mode and switch WiFi channels to detect in which channel the broadcast MAC address is received back. Once the channel is identified, we let the user device also send out its data packets and commands through the same broadcast method and let the IoT device do the same for data exchange. Because packet broadcasts can coexist with any ongoing WiFi sessions, the coexistence of immediate communication in proximity and communication through a control interface is guaranteed. We adopt this association-free communication as our default communication method for QuickTalk. 

\subsubsection{Validation}
To validate the proposed association-free communication, we perform the following experiment either at an indoor environment and at an outdoor environment. The experiment is to measure the RTT (Round-Trip Time) between the broadcast of a random payload of 24 bits at every 3 seconds from a user device to an IoT device and its reply of the same payload to the user device from the IoT device. For this experiment, we first do not use any application-level packet retransmission scheme in order to identify the pure performance of the proposed method\footnote{Although there is no application-level retransmission, link-level retransmissions from 802.11 standard may work.} and keep the signal strength between the user device and the IoT device between -30 and -60 dBm. 


Fig.~\ref{fig:echo} (a) shows the CDFs (Cumulative Density Functions) of RTT measured either at an indoor and an outdoor environment when there is no application-level packet retransmission. As shown in the figure, at an outdoor environment where there is almost no interfering WiFi signal, about 80\% of broadcast packets are successfully replied within in 0.01 seconds. At an indoor environment where we were able to scan about 30 interfering WiFi access points, within 0.01 seconds of RTT, about 60\% of broadcast packets are successfully replied. We further validate the performance of association-free communication when an application-level packet retransmission is implemented. Fig.~\ref{fig:echo} (b) shows the CDFs of RTT measured either at an indoor and an outdoor environment when we set the retransmission happens at every 0.25 seconds when the packet reception is not successful. As shown in the figure, we can observe substantial improvement in the reply rate as in both indoor and outdoor environments 
\textcolor{black}{more than 85 \% of broadcast packets are replied within 0.5 seconds}, confirming that the association-free communication is practically viable.

\section{IMPLEMENTATION}
In this section, we present implementation detail of a user device and an IoT device that use QuickTalk. As it is aforementioned, we exploit the Raspberry Pi 2 platform with IR circuits connected through GPIO (General Purpose Input/Output) in order to prototype both devices as shown in Fig.~\ref{fig:app}. \textcolor{black}{For the detailed analysis of our implementation, refer to our source code~\cite{QuickTalkCode}.} 

\savebox{\algleft}{
\begin{minipage}{.49\textwidth}
    \begin{algorithm}[H]
        \caption{\textcolor{black}{User device algorithms}}
        \small
        \label{alg:imp1}
        \begin{algorithmic}[1]
            \STATE //User device: IR service
            	
        	\STATE ($category_{\text{in}}$, $command_{\text{in}}$) = userinput()
        	\STATE sendEncodedIRMessage($category_{\text{in}}$)
            \STATE startWiFiservice()
            	        
        \end{algorithmic}
        
        \begin{algorithmic}[1]
            \STATE //User device: WiFi service
            \STATE ChannelArray[k] = getTopkRSSIs()
        	\IF {!channelDetected()}
                \STATE setRandomChannel(ChannelArray[])
            	\FORALL { ChannelArray[] }
            		\STATE setNextChannel()
            		\IF {($ch_{\text{IoT}}$, $\text{MAC}_{\text{user}}$)= receiveResponse()} 
            		    \BREAK
            		\ENDIF
            	\ENDFOR
        	\ENDIF
        	
        	\STATE broadcastMessage($command_{\text{in}}$, $ch_{\text{IoT}}$)
        	\STATE startPacketMonitor()
        	
        	\WHILE { TRUE }
        		\IF {$response_{\text{IoT}}$ = receiveResponse()}
                    \STATE displayResult($response_{\text{IoT}}$)
                    \BREAK
        		\ENDIF
        		\IF {needRETRANSMISSION()}
        			\STATE broadcastMessage($command_{\text{in}}$, $ch_{\text{IoT}}$)
        		\ENDIF
        	\ENDWHILE
            \STATE endPacketMonitor()
        \end{algorithmic}        
    \end{algorithm}%
\end{minipage}
}%
\savebox{\algright}{%
\begin{minipage}{.49\textwidth}
\begin{algorithm}[H]
    \caption{\textcolor{black}{IoT device algorithms}}
    \small
    \label{alg:imp2}
    \begin{algorithmic}[1]
        \STATE //IoT device: IR service          
        \STATE $\text{MAC}_\text{user}$ = parseIRMessage()
        
        \IF { !checkParity() or !checkCategory() }
            \STATE endProcedure()
        \ELSE
            \STATE startWiFiservice()
        \ENDIF
        

    \end{algorithmic}
    \begin{algorithmic}[1]  
        \STATE //IoT device: WiFi service        	
        \STATE startPacketMonitor()
        \WHILE {SWEEPING\_TIME\_OUT}
            \STATE broadcastMessage($\text{MAC}_{\text{user}}$, $\text{MAC}_{\text{IoT}}$)
            \STATE wait(BROADCAST\_INTERVAL)
            \IF {$command_{\text{user}}$ = ackReceived()}
                \STATE $sessionStart=$TRUE
                \BREAK
            \ENDIF
        \ENDWHILE
        
        \WHILE {TIME\_OUT $\&\&$ sessionStart}
            \IF {$command_{\text{user}}$ = commandReceived()}

            \STATE $response_{\text{out}}$ = processCommand($command_{\text{user}}$)
            \STATE broadcastMessage($response_{\text{out}}$)
            \ENDIF
        \ENDWHILE
        \STATE endPacketMonitor()
    \end{algorithmic}
    
\end{algorithm}
\end{minipage}}%

\noindent\usebox{\algleft}\hfill\usebox{\algright}%


\subsection{User Device Implementation}
Our implementation of user device consists of two parts: 1) user interface and 2) IR and WiFi services. 
The user interface is designed to get commands from a user either by clicking buttons, by typing commands, or by voice commanding. Our graphical user interface (GUI) is currently implemented by HTML and C++ using CGI (Common Gateway Interface)~\cite{robinson2004common} and the voice commanding function uses Google speech recognition APIs~\cite{GoogleSpeech}. For an immediate commanding to an IoT device, our user interface asks to provide the device type information along with a command. When the type information and the command are given, IR service first sends out the device type information and the MAC address of the user device as described in Section~\ref{sec:solution}, then WiFi service captures (i.e., monitors) the MAC address broadcasted by the triggered IoT device and delivers the command to that IoT device.
\textcolor{black}{Including this command delivery, all the following data exchanges through packet broadcasts use CoAP format provided Californium (Cf) JAVA library~\cite{kovatsch2014californium} for utilizing fundamental system of the IoT device.}

Fig.~\ref{fig:app} shows how the user interface implemented on a Raspberry Pi 2 device is presented to a user. IR and WiFi services therein are both implemented by C++ and use LIRC (Linux Infrared Remote Control) API for the operations of IR functions and socket API and MediaTek driver API for the packet broadcast and broadcast packet capture. The driver API is currently limited to the chipsets of MediaTek which includes MediaTek MT7601U (802.11 b/g/n) chipset that we connected to Raspberry Pi 2 devices through USB, but it is possible to extend the API for other WiFi chipsets. Upon capturing a broadcast packet, our WiFi service utilizes the packet capture library, \textit{libpcap}, to extract the contents from the broadcast packets and to detect the identity (i.e., WiFi MAC address) of the triggered IoT device. 

Because the IR module in the user device for QuickTalk is intentionally designed not to receive any information through IR communication, how to find the channel where the triggered IoT device broadcasts packets is a challenge. \textcolor{black}{To tackle this problem, our WiFi service is designed to randomly choose a channel among top $k$ channels using RSSI value and sweeps the channels one by one for 3 times. For this, we let the user device scan WiFi APs around and extract $k$ channels regularly in the background. For instance, because most APs of 2.4GHz in a building are positioned in about 3 non-overlapping channels to maximize their SNR performance, our method effectively narrows down the candidate channels of an IoT device in the proximity. Even in the worst of circumstances, having 3 iterations of sweeping for 11 channels is reliable to detect the channel where the IoT device is in. We describe the overall procedures that a user device goes through for commanding an IoT device as a pseudo code in Algorithm~\ref{alg:imp1}.}


\subsection{IoT Device Implementation}
Our implementation of an IoT device consists of two parts: 1) data processing service and 2) IR and WiFi services. 
The data processing service serves as the core of each IoT device, where the user command is processed and responded. The data processing service is also implemented by Californium library because it parses the CoAP queries and packages data in the CoAP format. Our implementation of the data processing service is general in that it can adapt to any form of IoT devices by a simple modification. Our current prototyping of IoT devices includes IoT bulb, IoT advertisement display, and IoT environmental sensor but not limited to those. The data processing service interacts with WiFi services through the loopback interface. 

The IR service in an IoT device is to receive an IR signal which includes the MAC address of a user device and the hierarchically designed device type filters and triggers the WiFi service to broadcast the MAC address when the device type filters match with the properties of the IoT device itself. For the reception of broadcast packets, the WiFi service in the IoT device also uses MediaTek driver API to \textcolor{black}{activate} the monitor mode (i.e., packet capture mode). 
We describe the overall procedures that an IoT device goes through as a pseudo code in Algorithm~\ref{alg:imp2}. \textcolor{black}{We have checked the power consumption of IR receiver at a Raspberry Pi 2 platform to see if it can critically reduce the operating duration of an IoT device. Our measurement shows that IR receiver consumes around 0.0158 W (5V, 3.15mA), which is very minor compared to the idle power consumption of Raspberry Pi 2 platform, 1.1 W (5V, 220mA) and its low-power modification, Pi Zero platform, 0.4 W (5V, 80mA).}

\subsection{Security Implementation}
The security problem related to controlling IoT devices in proximity is not our main focus. However, it is not very challenging to make QuickTalk communication secure and privacy protected. Suppose that our IR signal emitted from a user device contains a simple encryption key along with the ID information,  which can be authenticated by an IoT device. For a newly installed IoT device, the key can be given by the manufacturer while the key can be later updated by the user once QuickTalk communication is established. For the quick implementation of this idea, \textcolor{black}{we can adopt ECMV(Elliptic Curve Menezes-Vanstone)\cite{menezes1993elliptic} known as a low-complexity encipherment.} By doing so, it is possible to avoid accidental situations like IoT devices being controlled by unauthorized persons.

\section{EVALUATION}

\begin{figure}[t!]
	\centering
	\includegraphics[width=0.98\columnwidth]{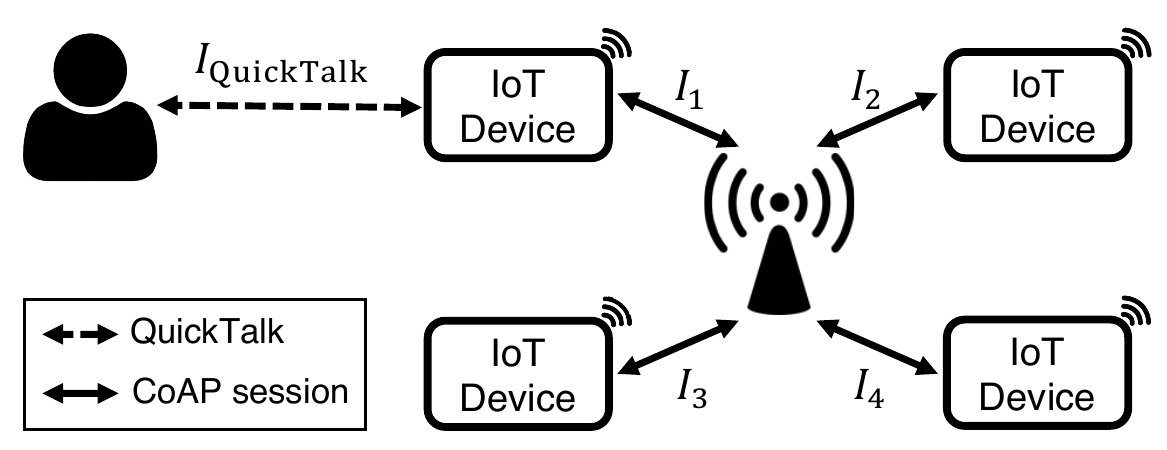}
	\caption{\textcolor{black}{The topology used for the evaluations of QuickTalk. $I_i$ denotes the packet arrival interval (second) of the CoAP communication associated with $i$-th IoT device,	whereas $I_{\text{QuickTalk}}$ stands for the packet arrival interval of QuickTalk between a user and an IoT device.}} 
    \label{fig:topology}
\end{figure}

We evaluate the performance of QuickTalk in a challenging situation where the candidate IoT devices to be controlled are previously registered to a control hub and are communicating with the hub through an WiFi access point as depicted in Fig.~\ref{fig:topology}. We assume that there are 4 registered IoT devices and a user device tries QuickTalk to one of those IoT devices. In such a situation, we first test if QuickTalk indeed enables an immediate commanding to an IoT device by measuring the end-to-end delay of QuickTalk, and then we further test how much performance degradation of the ongoing sessions between the IoT devices and the control hub experience when a user device communicates with one of the IoT devices through QuickTalk.

\begin{figure}[t!]
	\centering
	\includegraphics[width=0.98\columnwidth]{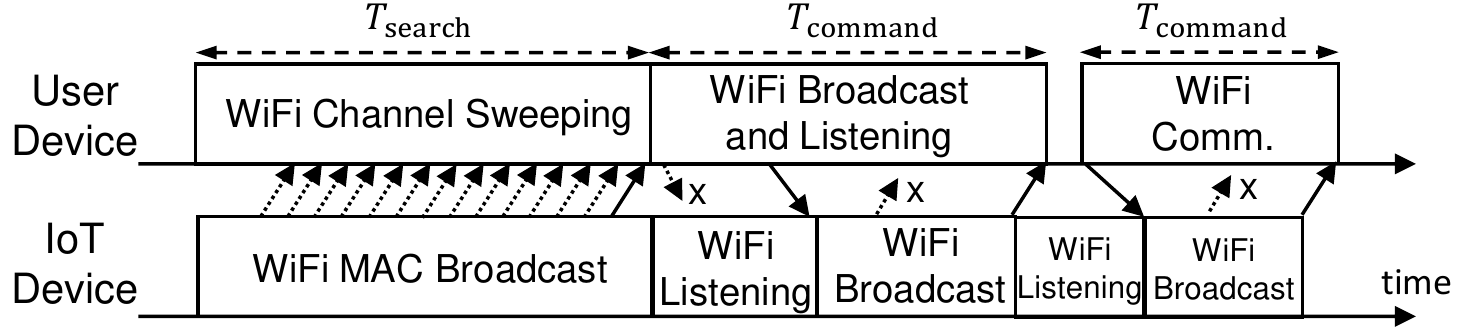}
	\caption{  \textcolor{black}{We draw the workflow of the WiFi service of QuickTalk.}
	We find that the end-to-end delay of QuickTalk is mainly affected by two major components: $T_{\text{search}}$ and $T_{\text{command}}$.}
    \label{fig:overall_flow}
\end{figure}

\begin{figure}[t!]
    \centering
    \subfloat[CDFs of $T_{\text{command}}$ with competing 4 CoAP sessions]
    {
        \includegraphics[width=0.98\columnwidth]{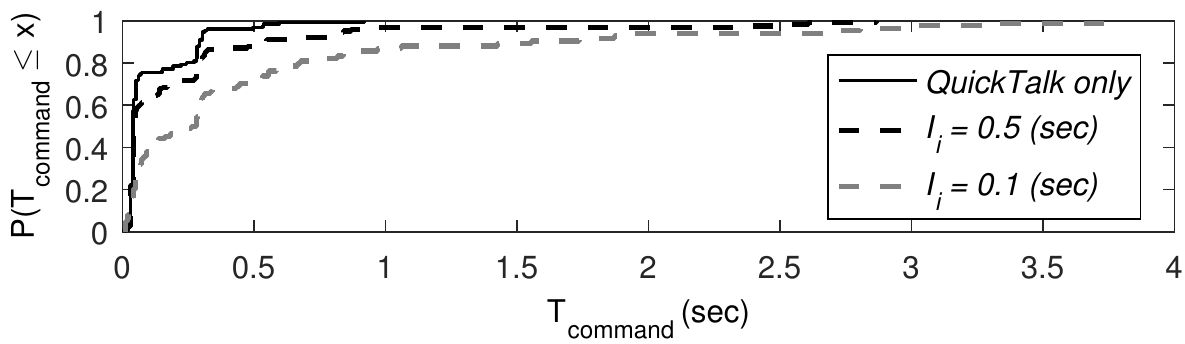}
    }
    \\
    \subfloat[\textcolor{black}{CDFs of $T_{\text{command}}$ by number of competing CoAP sessions that have interval of 0.1 seconds}]
    {
        \includegraphics[width=0.98\columnwidth]{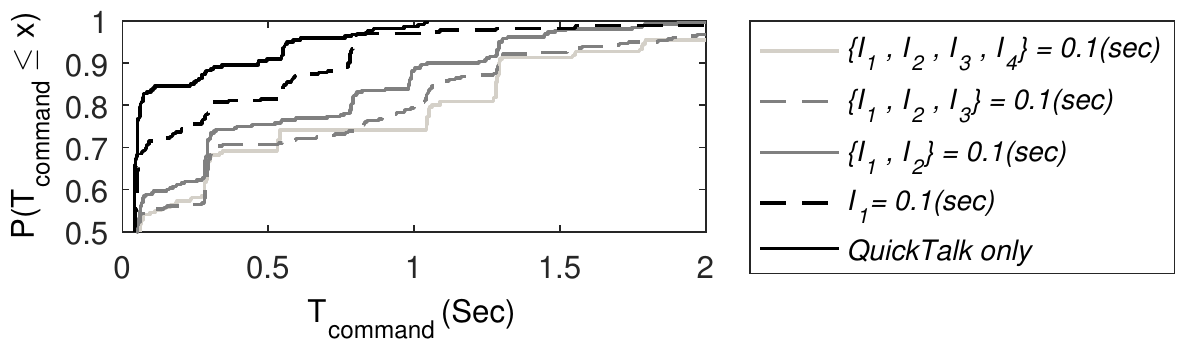}
    }
    \\
    \subfloat[CDFs of $T_{\text{search}}$ and the end-to-end delay]
    {
        \includegraphics[width=0.98\columnwidth]{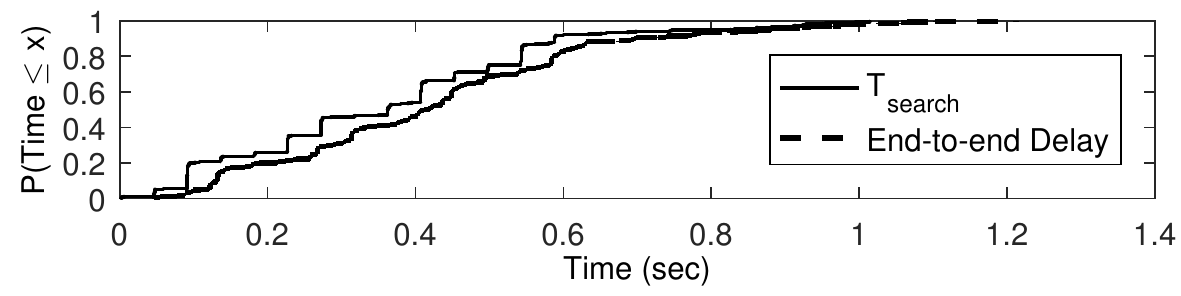}
    }
    \caption{    
    \textcolor{black}{CDFs of (a) $T_{\text{command}}$ when there are 4 competing CoAP sessions that 
    \textcolor{black}{have interval of 0.5 or 0.1 seconds} for each session, (b) $T_{\text{command}}$ when there are different number of competing CoAP sessions that have interval of 0.1 seconds for each session,
    and (c) $T_{\text{search}}$ and the end-to-end delay. The end-to-end delay of QuickTalk has its median at 0.414 seconds.}}
    \label{fig:e2edelay}
\end{figure}

\subsection{End-to-end delay of QuickTalk}
The end-to-end delay of QuickTalk from its IR signal transmission to the WiFi packet reception of the acknowledgment for an IoT command is mainly composed of two delay components: $T_{\text{search}}$ and $T_{\text{command}}$ as depicted in Fig.~\ref{fig:overall_flow}. Here, $T_{\text{search}}$ denotes the time duration of scanning channels to detect in which channel the IoT device triggered by an IR signal makes the broadcast. Since we set the channel switching duration as 40 ms in our experiment, the worst case of $T_{\text{search}}$ becomes 2 seconds given that our design takes two rounds of channel sweeping. The average delay for $T_{\text{search}}$ simply becomes the half of the worst case value. When the channel of interest is detected, the remaining delay is determined by $T_{\text{command}}$ where $T_{\text{command}}$ denotes the time duration between sending a command and receiving its acknowledgment through the broadcast channel of the WiFi interface. There are other delays in QuickTalk such as processing time for an IoT command, packet extraction time from a broadcast channel, and context switching delay from the IR service to the WiFi service, but we find that all such delays are on the scale of a few milliseconds in the processor of Raspberry Pi 2 (ARM Cortex A7, Quad-core, 900 MHz). Thus, we mainly focus on $T_{\text{search}}$ and $T_{\text{command}}$.  
\textcolor{black}{
Fig.~\ref{fig:e2edelay} (a) shows the CDF of $T_{\text{command}}$ when there is no ongoing CoAP session or when there are 4 ongoing sessions as depicted in Fig.~\ref{fig:topology}. For a QuickTalk session, we set the packet arrival interval (i.e., $I_{QuickTalk}$) by 5 seconds. For each ongoing session, we vary the 
packet arrival interval (i.e., $I_i$) by 0.5 and 0.1 seconds. As Fig.~\ref{fig:e2edelay} (a) confirms, QuickTalk mostly spends less than 0.5 seconds for $T_{\text{command}}$ when there is no ongoing CoAP session and experiences less than 1 second for $T_{\text{command}}$ at 80\% of the cases, when there are 4 busy CoAP sessions that exchange  
a packet at every 0.1 seconds. 
}
\textcolor{black}{
Fig.~\ref{fig:e2edelay} (b) shows the CDF of $T_{\text{command}}$ when there are different number of ongoing sessions as shown in Fig.~\ref{fig:topology}. For a QuickTalk session, we set the packet arrival interval (i.e., $I_{QuickTalk}$) by 5 seconds. We vary the number of ongoing sessions and we set the packet arrival interval (i.e., $I_i$) identically to be 0.1 seconds.
As Fig.~\ref{fig:e2edelay} (b) confirms, the CDF of $T_{\text{command}}$ decreases by the number of ongoing session. But, despite the channel is busy(4 ongoing sessions, $I_i = 0.1$), QuickTalk experiences less than 1 seconds for $T_{\text{command}}$ at 70\% of the cases.}
Fig.~\ref{fig:e2edelay} (c) shows the CDF of $T_{\text{search}}$ where is no ongoing CoAP session. Because we observe that having a number of ongoing CoAP sessions does not affect $T_{\text{search}}$, we only present the result with no ongoing CoAP session. As shown in Fig.~\ref{fig:e2edelay} (c), $T_{\text{search}}$ is relatively widely distributed from 0.04 seconds to 1.2 seconds since our channel sweeping algorithm naively starts from a randomly chosen channel among top 4 channels. Note that we can linearly speed up $T_{\text{search}}$ by reducing the channel switching delay, but for this, the hardware support is essential as the switching delay is currently bounded by the chipset delay. Also note that $T_{\text{search}}$ can be completely eliminated when a user device for QuickTalk is redesigned to receive IR signal from an IoT device regarding its current WiFi channel, but we consider that this is not user-friendly since this compels the user to keep its posture until the IR signal is successfully returned. 
     

 \begin{table}
 	\centering
 	\caption{\textcolor{black}{The throughput of an download session at an IoT device and the success rate of QuickTalk communication with that IoT device when the download session coexists with QuickTalk of various communication intervals.\label{table:ete}}}
     \begin{tabular}{|c|c|c|c|c|} \hline
        \specialcell{$I_{QuickTalk}$\\(sec)} & \specialcell{Download\\Only}& 10 & 5 & 3 \\ \hhline{|=|=|=|=|=|}
         \specialcell{Throughput\\(Mbps)}&18.54&17.53&15.75&14.40 \\ \hline
           \specialcell{Success\\Rate (\%)}&-&95.33\%&93.00\%&92.44\% \\ \hline
     \end{tabular}
\end{table}

The end-to-end delay of QuickTalk is also presented in Fig.~\ref{fig:e2edelay} (c) for the case where there is no ongoing CoAP session. As aforementioned, we find that the end-to-end delay is not much different from $T_{\text{search}} + T_{\text{broadcast}}$ and is upper limited by 1.2 seconds while its median is only about 0.414 seconds. Note that once the WiFi channel is detected, the end-to-end delay of QuickTalk approaches to $T_{\text{command}}$, which is roughly upper bounded by 1 second.


\subsection{Coexistence with Ongoing Sessions}
The coexistence of QuickTalk with ongoing communication sessions at an IoT device is an important matter given that there can be many IoT devices in practice, which are previously registered to control hubs but need to be immediately controlled by a user in proximity. To test the coexistence, we let one IoT device shown in Fig.~\ref{fig:topology} perform TCP-based file download and evaluate how much throughput degradation is observed when QuickTalk starts communicating with that IoT device for various commanding intervals\textcolor{black}{ (i.e. $I_{QuickTalk}$)}, 10, 5, and 3 seconds. Table~\ref{table:ete} shows the download average throughput, which is measured at the IoT device without and with QuickTalk. Table~\ref{table:ete} further shows the success rate of QuickTalk communication with an ongoing download session for various QuickTalk communication intervals. As the graphs show, the degradation of the throughput is limited to about 20\% when QuickTalk commands at every 3 seconds, compared to the download only case. 
\textcolor{black}{Also, the success rate of QuickTalk communication stays over 92\%} while the \textcolor{black}{interval of QuickTalk communication} varies from 10 seconds to 3 seconds. This experiment reveals that QuickTalk can reliably coexist with ongoing communication sessions in IoT devices.

\section{CONCLUSION AND FUTURE WORK}
\textcolor{black}{
In this work, we proposed QuickTalk, an association-free communication method for IoT devices in the proximity that is designed to enable intuitive, immediate and pinpointed communications with IoT devices around a user. QuickTalk works as a unified solution to the usability problems identified from various use cases in the reality: visitor, naming, and distance problems.
Our implementation of QuickTalk using Raspberry Pi 2 devices confirms that QuickTalk works reliably in the environments where multiple IoT devices run concurrent session and further shows that its end-to-end delay for delivering a command is reasonably low with the worst case bound of 2.5 seconds. We believe that QuickTalk that can be activated in every IoT device only by adding an IR receiver of a few cents can give a whole new user experience for IoT devices especially to non-tech savvy users. In the future design, we consider implementing the QuickTalk system on see-through devices allowing users to control the direction of pinpointing in conjunction with gaze tracking. 
}

\end{document}